# Photostability of gas- and solid-phase biomolecules within dense molecular clouds due to soft X-rays

S. Pilling[1]*, D. P. P. Andrade[1], E. M. do Nascimento[2], R. R. T. Marinho[2],
H. M. Boechat-Roberty[3], L. H. de Coutinho[3], G. G. B. de Souza[3], de Castilho R. B[3]
R. L. Cavasso-Filho[4] and A. F. Lago[4], A. N. de Brito[5]

[1] *Universidade do Vale do Paraíba (UNIVAP), São José dos Campos, SP, Brazil.*
[2] *Universidade Federal da Bahia (UFBA), Barreiras, BA, Brazil.*
[3] *Universidade Federal do Rio de Janeiro (UFRJ), Rio de Janeiro, RJ, Brazil.*
[4] *Universidade Federal do ABC (UFABC), Santo André, SP, Brazil.*
[5] *Laboratorio Nacional de Luz Síncrotron (LNLS), Campinas, SP, Brazil.*

**ABSTRACT**

An experimental photochemistry study involving gas- and solid-phase amino acids (glycine, DL-valine, DL-proline) and nucleobases (adenine and uracil) under soft X-rays was performed. The aim was to test the molecular stabilities of essential biomolecules against ionizing photon fields inside dense molecular clouds and protostellar disks analogs. In these environments, the main energy sources are the cosmic rays and soft X-rays. The measurements were taken at the Brazilian Synchrotron Light Laboratory (LNLS), employing 150 eV photons. In-situ sample analysis was performed by Time-of-flight mass spectrometer (TOF-MS) and Fourier transform infrared (FTIR) spectrometer, for gas- and solid- phase analysis, respectively. The half-life of solid phase amino acids, assumed to be present at grain mantles, is at least $3 \times 10^5$ years and $3 \times 10^8$ years inside dense molecular clouds and protoplanetary disks, respectively. We estimate that for gas-phase compounds these values increase one order of magnitude since the dissociation cross section of glycine is lower at gas-phase than at solid phase for the same photon energy. The half-life of solid phase nucleobases is about 2-3 orders of magnitude higher than found for amino acids. The results indicate that nucleobases are much more resistant to ionizing radiation than amino acids. We consider these implications for the survival and transfer of biomolecules in space environments.

## 1 INTRODUCTION

Prebiotic compounds such as amino acids and nucleobases (and related compounds) have been searched for in the interstellar medium/comets for at least the last 30 years (e.g., Brown et al. 1979, Simon & Simon 1973, Kuan et al. 2004). Some traces (upper limits) of the simplest amino acid, glycine ($NH_2CH_2COOH$), were observed in molecular clouds associated with star forming regions (Kuan et al. 2003a) and in the comet Hale-Bopp (Crovisier et al. 2004), but these identifications have not yet been verified (Snyder et al. 2005, Cunningham et al. 2007).

Despite the detection of amino acids in meteorites in the last 20 years (e.g., Engel et al. 1990; Pizzarello et al. 1991; Macko et al. 1997; Cronin 1998) their first conclusive detection from a comet was performed very recently by Elsila et al. (2009), using chromatographic and carbon isotopic analysis of samples collected from the comet 81P/Wild 2 and brought to Earth by the Stardust spacecraft. The authors have detected free glycine ($NH_2CH_2COOH$) and bound glycine precursors such as methylamine ($NH_2CH_3$) after acid-hydrolysis extraction of Stardust comet-exposed foils. These findings have enriched the understanding of comet chemistry and illustrated the potential delivery and survival of amino acids to the early Earth by comets, contributing to the prebiotic organic inventory from which life has supposedly emerged.

Despite no direct detection of nucleobases in comets or in molecular clouds, some of their precursor molecules such as HCN,

* E-mail: sergiopilling@pq.cnpq.br

pyridines, pyrimidines, and imidazole were reported in the Vega 1 flyby of comet Halley (Kissel & Krueger, 1987) and have been researched in the interstellar medium (Kuan et al. 2003b). The presence of other related ring-compounds such as benzene, naphthalene, and anthracene has been suggested in space environments such as dust shells of late stars, proto-planetary nebulaes (Cherchneff et al. 1992; Cernicharo et al. 2001a, 2001b), and cold molecular clouds (Iglesias-Groth et al. 2008, 2010).

The search for these prebiotic molecules in meteorites, on the other hand, has revealed a substantial number of proteinaceous and non-proteinaceous amino acids, up to 3 parts per million (ppm) (e.g., Cronin 1998 and references therein), and some purine and pyrimidine based nucleobases up to 1.3 ppm (e.g., Stocks & Schwartz 1981, Cronin 1998, Botta et al. 2007, Martins et al. 2008). This dichotomy between the carbonaceous chondrites meteorites and interstellar medium/comets remains a great mystery in the field of astrochemistry and in investigations concerning the origin of life.

This experimental photochemical study of gas phase and solid phase amino acids (glycine, DL-valine, DL-proline) and nucleobases (adenine and uracil) under a soft X-ray field is an attempt to test their stability against a high ionizing photon field which resembles those found in dense molecular clouds and protostellar disks. In these environments, the main energy sources are the cosmic rays and soft X-rays. Section 2, briefly presents the experimental setup. The results from the photodissociation of gas phase and solid amino acids and nucleobases in the presence of soft X-ray photons are presented and discussed in section 3. Some astrophysical implications, as well as an estimate for the half-lives of these molecules at possible astrophysical environments are provided and discussed in section 4. This section also presents a new methodology for the search of large organic molecules in astrophysical environments. Finally, in section 5, the final remarks and conclusions are given.

## 2 EXPERIMENTAL

The experiment was performed at the Brazilian Synchrotron Light Laboratory (LNLS), in Campinas, São Paulo, Brazil, employing 150 eV photons ($\sim 4\times10^{11}$ photons cm$^{-2}$ s$^{-1}$) from the toroidal grating monochromator (TGM) beam line (Fonseca et al. 1992, Cavasso Filho et al. 2005, Pilling et al. 2006, Pilling et al. 2007a). The photon flux at the sample was measured with a photosensitive diode (AXUV-100, IRD Inc.) mounted inside the experimental chamber and amounts to $\sim 4\times10^{11}$ photons cm$^{-2}$ s$^{-1}$. The entrance and exit slits of the beamline were in the range of 2 mm (wide slit mode) resulting in a bandwidth of about 1 eV. The molecule samples were commercially obtained from Sigma-Aldrich with purity better than 99.5%.

For the gas-phase experiments (glycine, adenine, and uracil), the solid samples were vaporized inside the vacuum chamber by a stainless steel sublimation apparatus located near the photochemistry location in which the soft X-ray beam perpendicularly intercepted the molecular beam (Lago et al. 2004, Pilling et al. 2007a). The produced ions were mass/charge analyzed by Time-of-Flight mass spectrometry (TOF-MS) employed in a photoelectron-photoion coincidence mode (Fig. 1a). TOF mass spectra were obtained using the correlation between one photoelectron and a photoion (Pilling et al. 2006). The ionized recoil fragments, resulting from the perpendicular interaction of the light with the gas sample, were accelerated by a two-stage electrical field and detected by two microchannel plate detectors in chevron configuration, after mass-to-charge (m/q) analysis by a TOF mass spectrometer. Due to the low vapor pressure of these compounds, it was necessary to use a heated source for sublimation, as mentioned before. The experiments were performed at about 25 °C above the temperature in which a sublimate was first noted. The base pressure in the vacuum chamber was in the $10^{-8}$ mbar range. The relative ionic yield for each ionic species (branching ratio) was determined directly from the mass spectra. The dissociation cross section were determined considering absolute absorption cross section values from the literature, as described elsewhere (Boechat-Roberty et al. 2005, Pilling et al 2007b, Boechat-Roberty et al. 2009).

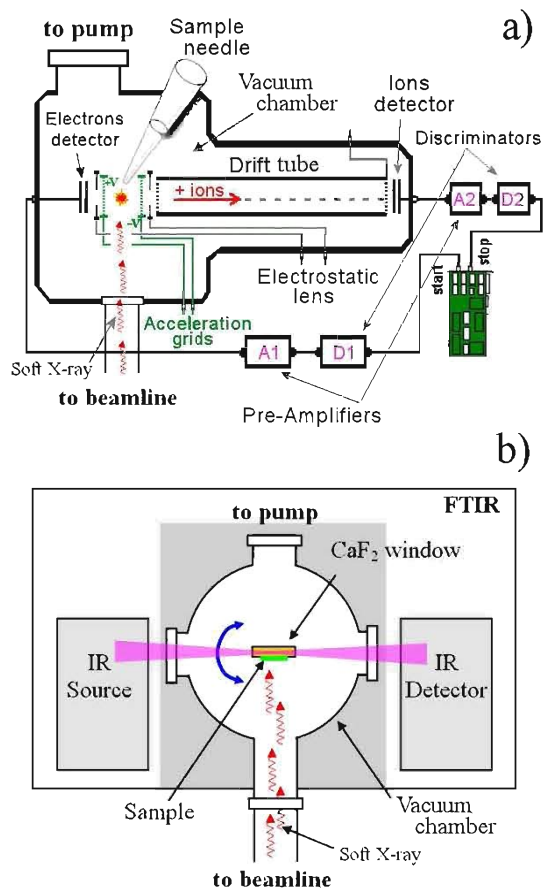

**Figure 1.** Schematic diagrams of the experimental setups. a) Gas-phase experimental setup: TOF mass spectrometer inside the experimental vacuum chamber and the associated electronics. a) Solid-phase experimental setup: Vacuum chamber with rotatable-axis sample holder coupled to an FTIR spectrometer.

For the solid phase samples (glycine, DL-proline, DL-valine, adenine, and uracil), in-situ analysis was performed by a Fourier transform infrared (FTIR) spectrometer (FTIR-400; JASCO Inc.) coupled to the experimental chamber. The spectra were obtained in the 3500 to 900 cm$^{-1}$ wavenumber range with a resolution of 1 cm$^{-1}$. 2 mm thick CaF$_2$ (water-insoluble) substrates were employed to hold the organic samples, being fully transparent to IR at the wavenumber range studied. For background correction, absorbance measurements was made before sample deposition. The analyzed samples were thin enough: i) to avoid saturation of the FTIR signal in transmission mode; and ii) to be fully crossed by the 150 eV photon beam.

The solid samples were diluted and deposited onto a CaF$_2$ substrate by drop casting following solvent evaporation at 50 °C

before their insertion into the vacuum chamber. Water was used as solvent for the amino acids. The sample concentration was about 10mg/mL and it was dropped about 1-5 microliters of the solution onto the $CaF_2$ substrate. An hydrophilic treatment on the substrate was employed before the sample deposition to make the substrate highly hygroscopic, which allowed the production of thin and homogeneous film on the surface. This treatment was performed by an $O_2$ plasma generator (Anatech Inc.). The $CaF_2$ dishes were submitted to a 600 mTorr of $O_2$ Plasma for about 30 minutes. After the treatment, the color of the substrates changed from transparent to light violet. No changes were observed in the IR spectra of the substrates.

Ethanol was used as the solvent for the water-insoluble compounds (nucleobases). About 50 mg of powder compounds were mixed with ethanol in a clean vessel and subjected to an ultrasonic bath for about 10 minutes. The sample concentration in each vessel was about 2-3 mg/mL and the volume, which was dropped into the $CaF_2$ substrate to produce the thin films, was about 30-50 microliters. Only the films with a high degree of homogeneity and with a thickness smaller than 3 microns were considered for the irradiation stage to guarantee an optical thin medium. The sample thicknesses were measured with a Dektak perfilometer (Vecco Inc.) and were of the order of 1-3 $\mu$m.

After the film preparation, the samples were placed into a vacuum chamber ($10^{-1}$ mbar) in order to test their stability under vacuum. No damage was observed. The samples were exposed to different soft X-ray doses (from 0.25 to 20 hrs). As previously pointed out, the *in-situ* sample analysis was performed by a Fourier transform infrared spectrometer coupled to the experimental chamber. The infrared beam from the FTIR and the synchrotron beam intercept the sample perpendicularly. The infrared transmission spectra were obtained by rotating the substrate/sample 90 degrees after each radiation dose. The measurements were done at room temperature. A schematic diagram of the experimental setup employed for the solid-phase experiments is shown in Fig. 1b. Infrared spectra of non-irradiated samples, taken at the beginning of the experiment and after several hours under vacuum ($10^5$ mbar), were compared. The comparison indicates that the solid sample did not evaporate from the substrate and the vacuum did not promote any spectral feature changes, as expected.

## 3 RESULTS AND DISCUSSION

### 3.1 Gas-phase experiments

Figure 2 presents the mass spectra of the fragments produced by the interaction of 150 eV photons on three gaseous samples: glycine ($NH_2CH_2COOH$), adenine ($C_5H_5N_5$), and uracil ($C_4H_4N_2O_2$). Each fragment peak in the spectra represents the opening of one photodissociation channel with the production of a cation (plus eventual release of neutral species). The integrated area of the given peak (ion) divided by the total area (total collected ions) times 100% is called the branching ratio (BR).[1] In the case of parental ions, the branching ratio are indicated by the label close to the rightmost peak (highest u/q) in each figure. The possible neutrals released in connection with the cationic species are marked by a blue text, between parenthesis, above several peaks.

---

[1] This calculation was performed at original Time-of-Flight mass spectra (counts × $\mu$s) and not at mass/charge spectra (counts × u/q). See details in Pilling 2006.

The fragmentation profile of gas-phase glycine is presented in Fig. 2a. The low branching ratio of the glycine molecular ion indicates that glycine is very sensitive to 150 eV photons. The most intense photodissociation channel contributes to about 23% of the opened channels and occurs via rupture of the C-C bond:

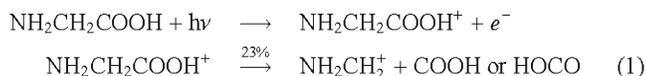

$$NH_2CH_2COOH + h\nu \longrightarrow NH_2CH_2COOH^+ + e^-$$
$$NH_2CH_2COOH^+ \xrightarrow{23\%} NH_2CH_2^+ + COOH \text{ or } HOCO \quad (1)$$

The alternative channel, in which the charge remains with carboxyl group or HOCO, is very unfavorable. Lattelais et al. (2010) have shown theoretically that the relative energies involving in the C-C rupture of ionized glycine with the release of $NH_2CH_2^+$ is about 19 kcal/mol, a value 3 times lower than the energy associated with the release of neutral $NH_2CH_2$, therefore being more favorable.

The second most intense dissociative channel occurs via formation of $HCNH^+$ (peak at 28 m/u) possibly together with neutral COOH + 2H or together with HCOOH + H. Considering these two most intense channels and peaks at 29 and 45 m/u, the amount of HCOOH or COOH (and its cation) released by 150 eV photons in glycine species is determined. For each hundred photodissociated glycine molecules in gas-phase, about 55 HCOOH or COOH (and its cation) molecules are produced.

The most probable (BR=27%) photodissociation channel in gas-phase adenine by soft X-ray occurs via with the release of a protonated hydrogen cyanide, $HCNH^+$:

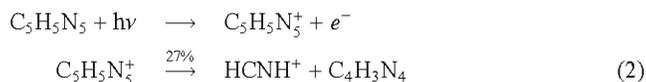

$$C_5H_5N_5 + h\nu \longrightarrow C_5H_5N_5^+ + e^-$$
$$C_5H_5N_5^+ \xrightarrow{27\%} HCNH^+ + C_4H_3N_4 \quad (2)$$

Hydrogen cyanide (HCN) or its cation are released in at least 7 dissociative channels. The amount of HCN or $HCN^+$ molecules in each dissociative channel and the respective branching ratio were used in order to determine the number of hydrogen cyanide molecules released by the dissociation of adenine using 150 eV photons. For each hundred photodissociated adenine molecules in gas-phase, about 80 HCN molecules (including its cation) are produced. This high production of HCN was also observed by Pilling et al. (2007a) in similar experiments using UV photon. Fig. 2b indicates that adenine is clearly less sensitive to radiation than glycine. The branching ratio for the adenine molecular ion is 8 times larger than the corresponding glycine molecular ion.

The dissociation profile of gas-phase uracil is shown in Fig. 2c. The branching ratio of uracil is half of the value determined for adenine and 4 times higher than the value determined for glycine. The most intense photodissociation channel observed occurs via liberation of protons. However, this peak must be interpreted with caution due to possible water contamination in the sample. The second most intense peak (BR ~ 13%) in the mass spectra occurs at mass/charge 28 being assigned as a protonated hydrogen cyanide ($HCNH^+$), together with possible release of a neutral isocyanic acid (HNCO) and ketenyl (HCCO) radical:

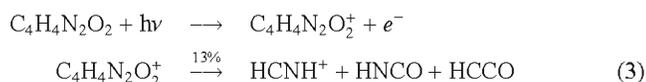

$$C_4H_4N_2O_2 + h\nu \longrightarrow C_4H_4N_2O_2^+ + e^-$$
$$C_4H_4N_2O_2^+ \xrightarrow{13\%} HCNH^+ + HNCO + HCCO \quad (3)$$

Contrary to what is observed in the photodissociation of adenine, the presence of HCN (or $HCN^+$) among the photodissociation channels is very unfavorable. Isocyanic acid (HNCO) is the most likely product formed from photodissociation of uracil by 150 eV photons. For each hundred photodissociated uracil molecules in gas-phase, about 40 HCNO molecules (including its cation) will be produced.

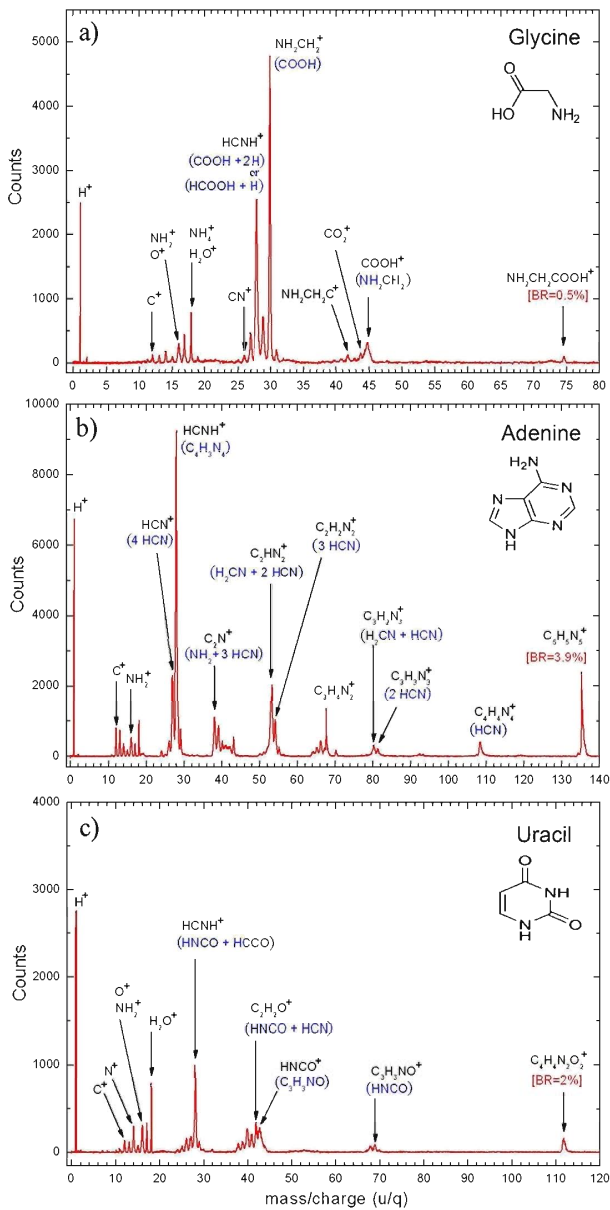

**Figure 2.** Mass spectra of the fragments produced by the interaction of 150 eV photons on gaseous samples. Branching ration (BR) for each parental ion, the rightmost peak, is indicated in each panel. The blue labels, between parenthesis, associated with some peaks indicate possible neutrals released in concert with the cationic species.

To put our data on an absolute scale, after subtraction of the linear background and false coincidences coming from aborted double and triple ionization (see details in Simon et al. 1991; Pilling et al. 2007c, 2007d), the contributions of all detected cationic fragments were summed and normalized to the photoabsorption cross sections, $\sigma_{abs}$, at 150 eV, taken from literature. The values employed were $\sigma_{abs} = 2.6 \times 10^{-18}$ cm$^2$, $1 \times 10^{-19}$ cm$^2$ and $2 \times 10^{-19}$ cm$^2$ for glycine (Kamohara et al. 2008), adenine (Moewes et al. 2004) and uracil (Moewes et al. 2004), respectively. We considered, as a first approximation, that the photoabsorption cross sections for solid phase (thin films) is very similar to the photoabsorption cross sections in the gas-phase, at the same photon energy. The photoabsorption cross sections determined at 150 eV represents roughly four times the photoabsorption cross sections at around 282 eV (extrapolated) for the studied species (see fig. 3 of Kamohara et al. 2008). Due to the lack of photoabsorption cross section measurements at 150 eV for uracil, it was adopted the averaged value between two other pyrimidine derivatives: cytosine, and thymine (Moewes et al. 2004).

Assuming a negligible fluorescence yield (due to the low carbon atomic number (Chen et al. 1981)) and anionic fragment production in the present photon energy range, it was adopted that all absorbed photons lead to cationic ionizing process. Therefore, the photodissociation cross section (also called dissociative single ionization cross section or photoionized dissociative cross section) of the gas phase studied molecules can be expressed by:

$$\sigma_{ph-d}(\text{gas}) = \sigma^+ \left(1 - \frac{BR_{M^+}}{100}\right) \quad (4)$$

where $BR_{M^+}$ represent the branching ratio of a given parental ion and $\sigma^+$ is the cross section for single ionized fragments (see description in Boechat-Roberty et al. 2005, Pilling et al. 2006). The determined values of $\sigma_{ph-d}(\text{gas})$ for the studied species due to the interaction with soft X-ray (150 eV) are given in Table 1.

### 3.2 Solid-phase experiments

The photostability of the solid phase amino acids (glycine, DL-valine, DL-proline) and nucleobases (adenine and uracil) due to the exposure of 150 eV photons as a function of time, is presented in Fig. 3 and Fig. 4, respectively. The spectra have been offset for better clarity. The upper spectrum of each panel represents the non-irradiated sample.

In the case of amino acids, the spectral features, in general, showed a clear decrease as a function of radiation dose. However, some functional groups seem to be more affected due to 150 eV photons than others, since individual infrared bands had different decreasing rates. The most sensitive group seems to be the C-H bands ($\sim 2500$ cm$^{-1}$) and the most resistant band is associated with asymmetrical stretching of $CO_2^-$ ($\sim 1600$ cm$^{-1}$). In the case of glycine and DL-proline, the absorption N-H band ($\sim 3200$ cm$^{-1}$) presents a large decreases as a function of irradiation time. The relationship between photostability and the molecular bonds has been carefully investigated and will be the subject of a future manuscript. No obvious new IR bands were observed. Since the experiments were performed at room temperature it is possible that the volatile species released from the dissociation (e.g. CO, $CO_2$, $H_2O$, $NH_3$, etc.) were desorbed from the sample. Next investigations performed at low temperature will focus on this subjection. For the peak attributions it was employed the works of Fischer et al. (2005), Suzuki et al. (1963), and Briget Mary et al. (2006).

As observed by Wilks et al. 2009, one of the effects of soft X-ray irradiation of organic compounds such as glycine, is the degradation of the carbonyl group which is highly sensitive to ionizing radiation. The authors have suggested from irradiation experiments at room temperature, also employing synchrotron soft X-rays on solid-phase glycine, that removal of an oxygen ion from the carbonyl group induces the formation of peptides in the sample.

On the other hand, the solid nucleobases were virtually not affected by the incoming soft X-ray radiation at 150 eV. The individual spectral features do not show any significant change during the irradiation phase. However, taking the spectrum as a whole, the integrated absorbance shows a small reduction as a function of radi-ation doses. This result clearly suggests that amino acids are more sensitive to soft X-rays photons than nucleobases.

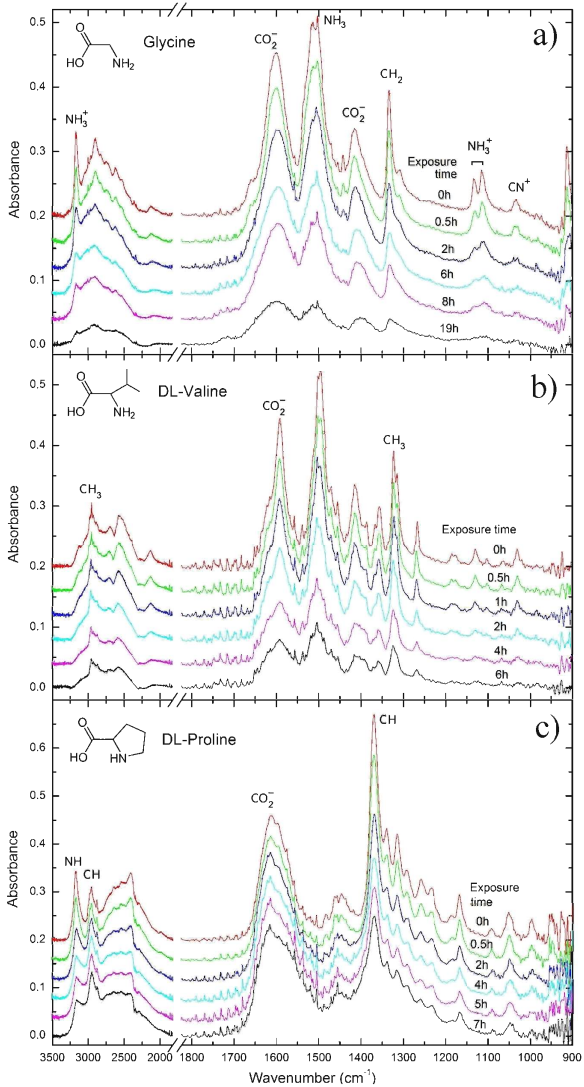

**Figure 3.** Photostability of the solid phase amino acids and nucleobases due to the exposure of 150 eV soft X-rays photons as a function of time.

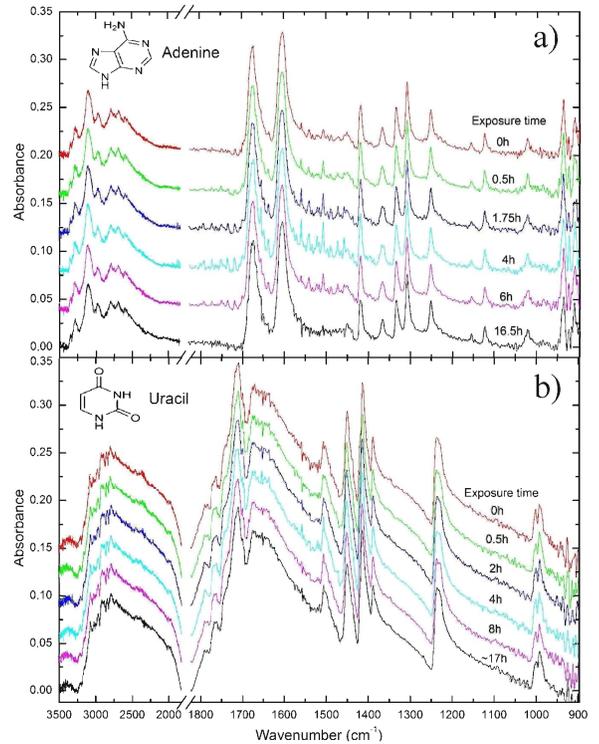

**Figure 4.** Photostability of the solid phase nucleobases due to the exposure of 150 eV soft X-rays photons as a function of time.

Assuming an optically thin medium and considering a constant 150 eV photon flux, the photodissociation rate, $k$, was determined from the evolution of the absorbance as a function of time. The dependence of the IR spectra integrated area (from 3500 to 900 cm$^{-1}$) as function of irradiation time, t, follows the Lambert-Beer formula: $\ln(A/A_0) = -kt$ where $A$ represents the total spectrum area at a given radiation dose and $A_0$ is the non-irradiated area. Figure 5 shows the integrated absorbance as a function of irradiation time for the studied solid samples. The horizontal doted line corresponds to the time in which the sample concentration is reduced to 0.5 of its initial value (half-life) due to irradiation in the laboratory. The values of the photodissociation rate for each molecular species are also given.

The most sensitive of the studied species was the DL-valine, which had the absorbance spectrum reduced to the half value after about 5.2 hours. The dotted lines are shown only to guide the eyes. For glycine, two sets of experimental data were taken (solid and open stars in Fig. 5) allowing the determination of the averaged half-life of 12 hours in the laboratory.

While in the gas-phase case, the cross section values were obtained from Eq. 4, the dissociation cross sections for the solid phase samples were derived from the measured photodissociation rate at a given photon flux. Considering that the photodissociation rate is generally given by the expression: $k = \int \sigma_{ph-d}(\varepsilon) F(\varepsilon) d\varepsilon$ where $\sigma_{ph-d}(\varepsilon)$ and $F(\varepsilon)$ are the photodissociation cross section and the photon flux as function of photon energy, respectively. Thus we can write, in a first approximation, that the dissociation cross section for solid samples can be given by the expression:

$$\sigma_{ph-d}(\text{solid}) \approx \frac{k_{150eV}(\text{solid})}{F_{150eV}} \quad (5)$$

where $k_{150eV}$, is the measured photodissociation rate due to 150 eV photons, and $F_{150eV} = 4 \times 10^{11}$ photons cm$^{-2}$ s$^{-1}$ is the measured photon flux at 150 eV at TGM beamline. This approximation is valid since the TGM beamline bandwidth is of the order of 1 eV (wide-slit mode plus off-focus mode).

The half-life, $t_{1/2}$, for the studied compounds in both gas and solid-phase experiments in the presence of 150 eV soft X-rays can be obtained directly by writing:

$$t_{1/2} = \frac{\ln 2}{k_{150eV}} \approx \frac{0.69}{F_{150eV} \sigma_{ph-d}} \quad (6)$$

which does not depend on the molecular number density. The determined values of $\sigma_{ph-d}$ and $t_{1/2}$ at gas and solid-phases for the studied species due to the presence of soft X-ray (150 eV) are given in Table 1.

**Table 1.** The determined values photodissociation cross section, $\sigma_{ph-d}$, and half-life, $t_{1/2}$, for the studied amino acids and nucleobases obtained by soft X-ray (150 eV). The possible values of half-life for these prebiotic molecules at different astrophysical environments are also given.

| | solid phase | | | | | gas phase | | | | |
|---|---|---|---|---|---|---|---|---|---|---|
| | $\sigma_{ph-d}$ | $t_{1/2}$ | | | | $\sigma_{ph-d}$ | $t_{1/2}$ | | | |
| Samples | $(10^{-18} cm^2)$ | Lab.[a] (hour) | XDRs[b] (year) | Dense Clouds[c] (year) | PPDs[d] (year) | $(10^{-18} cm^2)$ | Lab.[a] (hour) | XDRs[b] (year) | Dense Clouds[c] (year) | PPDs[d] (year) |
| Glycine | 40 | 12 | 70 | $7 \times 10^5$ | $7 \times 10^8$ | $2.6^e$ | 185 | $1 \times 10^3$ | $1 \times 10^7$ | $1 \times 10^{10}$ |
| Dl-Valine | 96 | 5.2 | 30 | $3 \times 10^5$ | $3 \times 10^8$ | - | - | - | - | - |
| DL-Proline | 13 | 38.5 | 200 | $2 \times 10^6$ | $2 \times 10^9$ | - | - | - | - | - |
| Adenine | 0.3 | ~1730 | $9 \times 10^3$ | $9 \times 10^7$ | $9 \times 10^{10}$ | $~0.3^f$ | ~1730 | $9 \times 10^3$ | $9 \times 10^7$ | $9 \times 10^{10}$ |
| Uracil | ~1.2 | ~330 | $2 \times 10^3$ | $2 \times 10^7$ | $2 \times 10^{10}$ | $~0.8^f$ | ~495 | $3 \times 10^3$ | $3 \times 10^7$ | $3 \times 10^{10}$ |

[a] $F_{150eV} = 4 \times 10^{11}$ photons $cm^{-2} s^{-1} \sim 10^3$ erg $cm^{-2} s^{-1}$ (off-focus TGM beamline)
[b] $F_{150eV} \sim 3 \times 10^{10}$ photons $cm^{-2} s^{-1}$ (typical XDR flux in molecular clouds; Meijerink & Spaans, 2005)
[c] $F_{150eV} \sim 3 \times 10^6$ photons $cm^{-2} s^{-1}$ (for AFGL 2591, at 200 AU from the X-ray source; Staüber et al. 2005)
[d] $F_{150eV} \sim 3 \times 10^3$ photons $cm^{-2} s^{-1}$ (for typical T Tauri disks at 25 AU into the disk from the central star; Agúndez et al. 2008 and Nomura et al. 2007)
[e] Absorption cross section taken from Kamohara et al. 2008.
[f] Absorption cross section taken from Moewes et al. 2004.

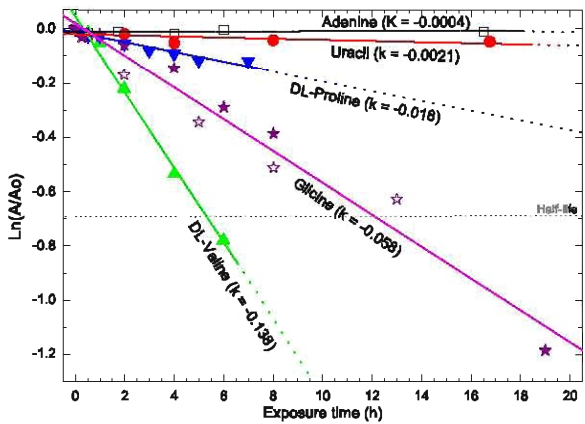

**Figure 5.** Integrated absorbance spectra of the solid phase samples as a function of irradiation time. For each compound, the photodissociation rate, $k$, is also indicated.

## 4 ASTROPHYSICAL IMPLICATIONS

The presence of all prebiotic compounds studied in this work was detected in carbonaceous chondrites meteorites (e.g. Cronin 1998, Martins et al. 2008 and references therein) which reveal the pristine chemistry of the early solar nebula. Therefore, they are assumed to be produced in extraterrestrial environments. However, the exact formation location, either in dense clouds or in proto-planetary disks, is not known. In this section, we will hypothetically assume that these species could be formed at dense clouds and throughout cloud evolution become members of the chemical inventory of molecular cloud cores and proto-planetary disks (PPDs).

The outer parts of dense molecular clouds are highly exposed to ultraviolet radiation (and X-rays) and the molecules are fully dissociated/ionized (HII regions). By moving deeper into the cloud ($0.1 > A_v > 5$; Agúndez et al. 2008), the UV may now promote the dissociation/ionization of CO, water, and simple organic species in the gas phase and also heat and evaporate some molecules trapped on grains (photodissociation regions - PDRs). At the intermediate regions inside the clouds ($>10^{22}$ cm$^{-2}$ or $A_v > 5$), virtually all the UV photons are absorbed and only the X-ray from nearby stars (Maloney et al. 1996) may interact with the gas and the dust grains, leading to different photochemical reaction pathways, which may increase the molecular complexity.

Depending on the number density of the molecular cloud and its gradient along the cloud, the X-ray photochemistry regions (X-ray-dominated photodissociation regions XDRs) could be extended deeper into the clouds until the visual extinction reaches a value over $10^{25}$ cm$^{-2}$ ($A_v \gtrsim 5000$) (Agúndez et al. 2008). At this point X-rays are severely attenuated by molecular hydrogen or absorbed/scattered by the grains.

As discussed by Meijerink & Spaans (2005), the typical XDR energy flux in molecular clouds is about $\sim 10$ erg cm$^{-2} s^{-1}$ which decreases as visual extinction increases within the clouds. Considering a constant photon flux inside the wavelength range employed by the authors (0.1 to 10 keV), the photon average flux at 150 eV is estimated to be $\sim 3 \times 10^{10}$ photons cm$^{-2} s^{-1}$.

Following Staüber et al. (2005), the radiative flux in the X-ray domain inside the dense molecular cloud AFGL 2591 can be described by a blackbody with temperature $T_x \gtrsim 3 \times 10^7$ K resulting a typical X-ray luminosity $L_x \gtrsim 10^{31}$ erg s$^{-1}$. Taking into account only the thermal emission, we have estimated the photon flux at 150 eV at the distance of 200 AU, from the central source, the lower distance used in the AFGL 2591 X-ray chemistry models by Staüber et al. (2005). Employing a similar methodology described at Pilling et al. (2007d), the value obtained for 150 eV photons was $3 \times 10^6$ photons cm$^{-2} s^{-1}$. For deeper distances into the dense cloud, as well for higher soft X-ray photon energies, the flux is even lower.

In an attempt to define a region where a significant flux of soft X-rays photons can penetrate in contrast with the UV photons, we adopted a region with a visual extinction of about $A_v \sim 4$. We used the standard relation $A_v = N_H$ (cm$^{-2}$)/ $1.87 \times 10^{21}$ (Bohlin et al. 1978), and an averaged gas density inside PPD of about $2 \times 10^8$ cm$^{-3}$ (Agúndez et al. 2008) to determine an effective region in which soft X-ray may rule out the photoionization and photodissociation processes in gas and on grain surfaces. This region is about 25 AU from the star ($A_v \sim 4$).

Nomura et al. (2007) have studied a typical Class-I T Tauri star TW Hya at soft X-rays and X-rays. The authors have calculated, for the 150 eV photon energy, a photon flux of $\sim 10^{-4}$ photons cm$^{-2} s^{-1}$ which represents about ~ $10^{33}$ photons cm$^{-2}$ s$^{-1}$ at the star (taking into an account that TW Hya is about 56 pc ~ $3 \times 10^{18}$ cm from us). Considering the very low opacity ($\tau$ <1) for soft X-ray up to 25 AU from the center, the estimated flux at 150 eV for Typical T Tauri stars at 25 AU ($A_v$ ~ 4) into the disk is about $F_{150eV} \gtrsim 3 \times 10^3$ photons cm$^{-2}$ s$^{-1}$ which represents an energy flux of ~ $10^{-5}$ erg cm$^{-2}$ s$^{-1}$.

The estimated values for the half-lives for these important prebiotic molecules at different astrophysical environments are presented in Table 1. These values were calculated using the experimental photodissociation cross section (also presented in this table) and the soft X-ray flux at several sources taken from literature (typical X-ray photodissociation regions-XDR (Meijerink & Spaans (2005)), dense molecular clouds (Stauber et al. 2005), proto-planetary disks (Agúndez et al. 2008). The most sensitive species is the DL-valine, which is predicted to survive only about 30 years under direct X-ray exposure at a typical XDR. Whereas, the adenine molecule could survive almost $10^4$ years on the same X-ray photon flux. However, all the species seem to survive more than $10^5$ years into dense clouds and more than $10^8$ years into protoplanetary disks. Despite this, the results seem to give reasonable values for half-lives in these regions, in comparison to life-times of the astrophysical environments. More observational data at soft X-ray range (or modeling) are needed to increase the reliability of the calculated half-lives.

The results presented in Table 1 indicate that solid-phase amino acids can survive at least $3 \times 10^5$ years and $3 \times 10^8$ years in dense molecular clouds and protoplanetary disks, respectively. For the nucleobases the photostability is even higher, being about 2-3 orders of magnitude higher as compared to the most radiation sensitive amino acid. This high degree of survival may be attributed to the low photodissociation cross section of these molecules in the soft X-ray range combined with the protection promoted by the dust that reduces the X-ray field within denser regions. The half-life of glycine in gas-phase was found to be about 15 times higher (due to the smaller photodissociation cross section) than the determined value in the solid phase. With adenine and uracil no significant changes were observed in half-lives between gas- and solid-phases.

The nucleobases half-lives were up to two orders of magnitude higher than for the amino acids which could be attributed to the presence of molecular resonance (aromaticity). Following the molecular orbital calculations performed by Pullman & Pullman (1960; 1962), purines and pyrimidines have the large resonance energy which confer a high radiation stability degree upon them.

As indicated by Allamandola et al. (1989) the nucleobases are able to form a stable cation in the gas phase. Because polycyclic aromatic hydrocarbons (PAHs) and polycyclic aromatic nitrogen-rich hydrocarbons (PANHs) have the same capability and seem to be ubiquitous in the ISM, it is not unreasonable to predict that aromatic nucleic acid bases could survive in the interplanetary and interstellar media. Recently, Iglesias-Groth et al. (2008, 2010) detected a strong band of naphthalene cation ($C_{10}H_8^+$) and anthracene cation ($C_{14}H_{10}^+$), the two and three ring simplest linear PAHs, respectively, in the line of sight of star Cernis 52, a likely member of the very young star cluster IC 348, which is probably associated with cold absorbing material in a intervening molecular cloud of the Perseus star forming region. These compounds have column density of ~ $1.1 \times 10^{13}$ cm$^{-2}$ implying that about 0.02 % of the carbon in the cloud could be found in the form of these two aromatic compounds.

Previous studies on gas-phase also showed that even at the vacuum ultraviolet (VUV) region amino acids are destroyed more effectively by stellar radiation than the nucleobases (e.g. Pilling et al. 2007a, Pilling et al. 2008, Lago et al. 2004, Coutinho et al. 2005, Marinho et al. 2006).

**Table 2.** Comparison between the half-lives values of solid-phase amino acids and nucleobases in dense clouds due to UV cosmic-ray induced flux and soft X-ray flux.

| Samples | Half-life in Dense Clouds (Myr) | |
|---|---|---|
| | UV[a] | Soft X-rays[b] |
| Glycine | 1.84[c] | 0.7[d] |
| Dl-Valine | - | 0.3 |
| DL-Proline | - | 2 |
| Adenine | 8.27[c] | 90[d] |
| Uracil | 2.03[c] | 20[d] |

[a] Assuming a cosmic-ray induced UV flux of $10^3$ photons cm$^{-2}$ s$^{-1}$ (Prasad & Tarafdar 1983)
[b] Assuming a 150 eV photon flux of $3 \times 10^6$ photon cm$^{-2}$ s$^{-1}$ (for AFGL 2591, at 200 AU from the X-ray source; Staüber et al. 2005)
[c] Pure compounds at 12 K in Argon matrix; Peeters et al. 2003.
[d] Pure compounds at room temperature; this work.

Table 2 presents a comparison between the half-lives of solid-phase amino acids and nucleobases in dense clouds due to UV cosmic-ray induced flux (from the irradiation of ices at low temperature; Peeters et al. 2003) and soft X-ray flux (150 eV; this work). The high photostability of nucleobases is also observed under UV radiation. The half-life of glycine measured from the irradiation of pure compounds in Ar matrix at 12 K employing a hydrogen lamp (Peeters et al. 2003), is about 2.5 higher than the measured employing 150 eV soft X-rays. However, in the case of nucleobases this difference reaches one order of magnitude. The high photostability of frozen amino acids due to UV was also observed by Ehrenfreund et al. 2001 in similar matrix spectroscopy experiments employing different gases.

A radiation-induced decomposition experiment at room temperature performed by Wilks et al. (2009), employing soft X-ray around 290 eV (C K-edge) and 400 eV (N K-edge) on glycine powder, shows that major effect of radiation is the fragmentation of molecule at the carbonyl sites. The authors also observed a peptide formation (e.g. diglycine, triglycine) under irradiation associated with the removal of an oxygen ion from the carbonyl group. Due to the low spectral resolution of the current experiment no evidence of such peptide formation is given. As observed by Wilks et al. (2009), $NH_2CH$ and $NHCH_2$ contributes with about 20% of the photodissociation products. This species were also observed among the fragments produced by the photodissociation of glycine at gas-phase (small peak at 29 m/u on Fig.2a). A possible formation route of these species in solid phase could be attributed with the deprotonation of $NH_2CH_2^+$ which is the most abundant fragment of glycine at gas-phase. Future experiments employing isotopic label and low temperature will helps to elucidate this issue.

Recently, Lattelais et al. (2010) investigated the survival of pure glycine ice and glycine:$H_2O$ ice at 30 K after the irradiation by soft X-rays photons with energies around 535 eV (O K-edge). They observed that the presence of water does not enhance or protect glycine from photodecomposition and the role of trapped OH radical in the destruction of glycine is negligible. Employing their data[2], we estimated the value for the dissociation cross section of frozen glycine (with or without water) by soft X-rays, about $4 \times 10^{-20}$ cm$^2$. This value is three order of magnitude lower than the current value obtained for solid glycine at room temperature irradiated by 150 eV photons (see Table 1). Considering that this difference is mainly ruled by temperature, this suggests that the half-life of glycine at astrophysical scenarios can be even higher than the listed values given in Table 1, if this compound were found frozen at interstellar grains. This results corroborate with the scenario that during planetary formation (and after) molecules such these, which are trapped in and on dust grains, meteoroids, and comets, could survive long enough to ionizing radiation field (X-rays) to be delivered to the planets/moons possibly allowing prebiotic chemistry occurs.

As discussed before the gas-phase investigation indicated some compounds that are associated with the photodissociation of its parental species. For example, for each 100 photodissociated glycine by soft X-rays, about 55 molecules of HCOOH (recombination product) plus COOH molecules will be produced. In the case of the photodissociation of 100 adenine molecules and 100 uracil molecules, 80 HCN (and its cation) and 40 HNCO (and its cation) will be produced, respectively. Moreover, although adenine and uracil have not been yet detected on extraterrestrial environments, at least one of their abundant fragments from photodissociation by soft X-rays, the HCNH$^+$ ion, has been extensively found in the interstellar medium (Ziurys et al. 1986; Schilke et al. 1991), comets (Ziurys et al. 1999), and planetary atmospheres such as in Titan (Vacher et al. 2000, Petrie 2001). For the gas-phase experiments, these abundant daughter fragments represent a typical signature of each compound processed by the studied soft X-ray. Additionally, is also reasonable to expect that, peculiar fragments may be used as a signature of their parental species, for example, a backbone fragment produced by the hydrogen's stripping resulting from the photodissociation of large molecule.

A combination between the experimental branching ratio of these daughter fragments and their measured column density at interstellar sources from radio-astronomical observation may be used to put some upper limits values for the abundance of their parental species in gas phase. However, since these compounds can react to form new products or they can also be produced from different parental species, only upper limits for abundances can be estimated. Moreover, it should be noted that these molecular fragments in regions associated with ices and any energetic processes may also (and likely do) form other larger organic compounds (see Gibb et al. 2004; Bernstein et al. 2002; Muñoz Caro et al. 2002). Therefore, the observation of daughter species in this line of sights (for example, by infrared spectroscopy) may not be delimiting for the presence of larger organic compounds in solid-phase (ices).

## 5   CONCLUSION

We have presented experimental studies on the interaction of soft X-rays on gas-phase and solid-phase amino acids and nucleobases in an attempt to verify if these molecules can survive long enough to be observed or even to be found in meteorites. Measurements were performed employing 150 eV photons under high vacuum conditions at the Brazilian Synchrotron Light Laboratory (LNLS).

[2] Considering both Fig. 2 and the photon flux information at the footnote of page 4 of Lattelais et al. 2010.

The produced ions from the gas-phase experiments (glycine, adenine, and uracil) were mass/charge analyzed by Time-of-Flight spectrometer (TOF-MS). The analysis of solid phase samples (glycine, DL-proline, DL-valine, adenine and uracil) were performed by a Fourier transform infrared (FTIR) spectrometer coupled to the experimental chamber. Photodissociation cross sections and half-lives were determined and extrapolated to astrophysical environments. Our main results and conclusion are the following:

(i) The results show that the gaseous amino acids are largely destroyed by soft X-rays while the nucleobases have a survivability (branching ratio of parental ion) of a few percents. As expected, a similar behavior was also observed in the case of solid samples.

(ii) The fragmentation profile on gas phase reveals that the most probable products from the photodissociation of glycine, adenine, and uracil by 150 eV photons are HCOOH or COOH (or/and its cation), HCN (or/and its cation), and HNCO (or/and its cation), respectively.

(iii) A comparison between gas- and solid-phase bio molecules reveals that the dissociation cross section of solid glycine is about 10 times higher than in the gas-phase. This suggests that in addition to radiation protection, the presence of a catalytic surface such as an interstellar or interplanetary grain surface can decrease the half-life of photosensitive compound due to an increase in the number of the dissociation channels (low-activation barrier channels or free-activation channels). In nucleobases, the dissociation cross section to X-ray photons are roughly the same at both solid and gas-phase not showing any excess dissociation due to catalytic surface.

(iv) The nucleobases photostability is up to two orders of magnitude higher than for the amino acids. The condensed amino acids can survive at least $\sim 3 \times 10^5$ years and $\sim 3 \times 10^8$ years in dense molecular cloud (MC) and PPD, respectively. For nucleobases the determined half-life is of the order or even grater than MC and PPD lifetime.

The results obtained in the present work corroborate with the scenario that during planetary formation (and after) these molecules, which are trapped in and on dust grains, meteoroids, and comets, could survive long enough to ionizing radiation field (X-rays) to be delivered to the planets/moons possibly allowing prebiotic chemistry in such environments where water was also found in liquid state. These results lead us to ask an interesting question. Why don't we find nucleobases in radioastronomical observations of cometary/molecular clouds since they are more resistant to stellar ultraviolet radiation than the detected amino acid (e.g., glycine)? Probably, the answer may be associated with the efficiencies of formation pathways rather than with the detection limits and thus more theoretical and laboratory studies of this subject are need.

## ACKNOWLEDGMENTS

The authors would like to thank the staff of the Brazilian Synchrotron Facility (LNLS) for their valuable help during the course of the experiments. We are particularly grateful to Dr. Avran Slovic, Dr. Flavio Vicentin, Dr. Guinter Kelerman, Dr. Paulo de Tarso, Dr. Maria Helena and Dr. Angelo Gobbi. We also express our gratitude to the vacuum office team, the mechanical office team and the thin films laboratory team. We also thanks Ms. Alene Rangel for the English revision of this manuscript. This work was supported by LNLS, CNPq, FAPERJ and FAPESP.

This paper has been typeset from a T<sub>E</sub>X/ L<sup>A</sup>T<sub>E</sub>X file prepared by the author.